\documentclass[prb,twocolumn,superscriptaddress,showpacs,amsmath,amssymb]{revtex4}

\usepackage{graphicx}
\graphicspath{{fig/}}
\usepackage{dcolumn}
\usepackage{bm}

\begin{document}

\preprint{APS/123-QED}

\title{
Motion of a free-standing graphene sheet induced by a collision with an argon nanocluster \\
: Analyses of the deflection and the heat-up of the graphene
}

\author{Kuniyasu Saitoh}
\affiliation{%
Yukawa Institute for Theoretical Physics, Kyoto University, Sakyo-ku, Kyoto, Japan
}%
\affiliation{%
Department of Mathematics, University of Leicester, Leicester LE1 7RH, United Kingdom
}%
\author{Hisao Hayakawa}%
\affiliation{%
Yukawa Institute for Theoretical Physics, Kyoto University, Sakyo-ku, Kyoto, Japan
}%

\date{\today}

\begin{abstract}
Nanocluster impact on a free-standing graphene is performed by the molecular dynamics simulation, and the dynamical motion of the free-standing graphene is investigated.
The graphene is bended by the incident nanocluster, and a transverse deflection wave isotropically propagated in the graphene is observed.
We find that the time evolution of the deflection is semi-quantitatively described by the linear theory of elasticity.
We also analyze the time evolution of the temperature profile of the graphene,
and the analysis based on the least dissipation principle reproduces the result in the early stage of impact.
\end{abstract}
\maketitle

\section{introduction}
Graphene is a two-dimensional (~2D~)  atomic layer of carbon atoms on a honeycomb lattice.
Recent remarkable experimental techniques have made it possible to observe the motion of a free-standing or suspended graphene sheet \cite{susp-g,free-g}.
Because electrons in a graphene can travel sub-micrometer distances without scattering, the study of graphene is active to make nanoscale electronic devices \cite{elec-g}.
Graphene can be wrapped up into fullerenes and rolled into carbon nanotubes, and thus it is the most fundamental structure of nano-carbon materials \cite{revw-g}.
Such flexibility of graphene encourages many researchers to investigate its mechanical properties.
A recent experiment has detected the mechanical vibrations of suspended graphene sheets activated by radio frequency voltages,
and has observed vibration eigenmodes which are not predicted by the elastic beam theory \cite{mech-g}.
In contrast to the electrical activations of graphene,
it is also possible to activate the motion of graphene by nanocluster impact \cite{mech-sg4}.
The nanocluster impact can generate high pressure in localized areas of graphene,
and it is an appropriate method to verify the elastic theory for the plate deflected by the concentrated force.
In addition, nanocluster impact is also important for manufacturing nanoscale electronic devices on a substrate \cite{dep1,dep2,dep3,dep4,ptp}.
Therefore, it is necessary \textit{to understand the motion of the graphene induced by a collision with nanocluster}
in order to verify the elastic theory and to aim to construct the nanoscale electronic devices on a graphene sheet.
However, there are a few studies which investigate the time evolution of the local deformation of the graphene deflected by the nanocluster impact.
In this paper, we perform the molecular dynamics (~MD~) simulation to investigate the time evolution of the deformation
of a free-standing graphene sheet deflected by a collision with an argon nanocluster.
We find that analytic solutions of the elastic plate well reproduce the results of our MD simulation.
We also analyze the time evolution of the temperature profile of the graphene sheet.

The organization of this paper is as follows.
In Section \ref{sec:setting}, we introduce our numerical model of the nanocluster impact on a graphene sheet.
Section \ref{sec:results} consists of three subsections.
In Section \ref{sub:observe}, we show the time evolution of the deflection of the graphene.
In Section \ref{sub:deflect}, we analyze the time evolution of the deflection.
In Section \ref{sub:temp}, we analyze the heat-up of the graphene after the impact.
We discuss our results in Section \ref{sec:disc} and conclude in Section \ref{sec:conc}.

\section{molecular dynamics simulation of the impact
\label{sec:setting}}
\begin{figure}
\includegraphics[width = 9 cm]{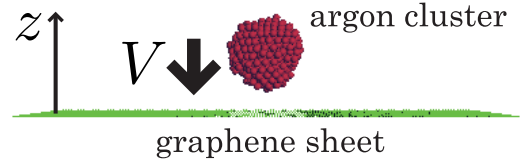}%
\caption{(Color online)
A snapshot of impact of an argon cluster on a free-standing graphene sheet.
The incident cluster contains $500$ argon atoms.
The graphene sheet contains $16032$ carbon atoms on a honeycomb lattice.
\label{fig:setting}}
\end{figure}
To study the dynamical motion of the graphene induced by a collision with an argon cluster, we perform the MD simulation.
We adopt the Lennard-Jones (~LJ~) potential $\phi(u) = 4\epsilon \left[ (\sigma/u)^{12} - (\sigma/u)^6 \right]$ for the interaction between two argon atoms
with the distance $u$ between two argon atoms, where we use the LJ parameters~\cite{LJ} $\epsilon=1.03 \times 10^{-2}$~$(\mathrm{eV})$ and $\sigma=0.340$~$(\mathrm{nm})$.
We also adopt LJ potential for the interaction between an argon atom and a carbon atom,
where we use the cross parameters of LJ potential $\epsilon_{\mathrm{int}}$ and $\sigma_{\mathrm{int}}$,
which are defined by the Lorentz-Berthelot rule as $\epsilon_{\mathrm{int}}=\sqrt{\epsilon \epsilon'}$ and $\sigma_{\mathrm{int}}=(\sigma+\sigma')/2$, respectively.
Here, $\epsilon' = 2.40 \times 10^{-3}$~$(\mathrm{eV})$ and $\sigma' = 0.335$~$(\mathrm{nm})$ are the LJ parameters for carbon \cite{modLJ1,modLJ2}.
Finally, we adopt the Brenner potential, which is widely used for simulations of a graphene and a carbon nanotube, for the interaction between two carbon atoms \cite{bren}.

Figure \ref{fig:setting} displays a snapshot of our impact simulation.
The graphene involves $16032$ carbon atoms on a honeycomb lattice.
The bond length of the graphene is $0.142\mathrm{nm}$ and the length of one edge is approximately equal to $20\mathrm{nm}$.
The carbon atoms on the edges parallel to the $x$-axis are arranged in armchair geometries,
and the carbon atoms on the edges parallel to the $y$-axis are arranged in zigzag geometries \cite{mech-sg1,mech-sg2}.
The boundary conditions of the four edges of the graphene are free, and the initial temperature of the graphene is $1.2 \mathrm{K}$.
The cluster containing $500$ argon atoms is made from argon gas by the temperature quench method \cite{quench1,quench2}.
At first, we prepare $500$ argon atoms in a periodic box and equilibrate at $119.6 \mathrm{K}$ with the number density $1.27 \mathrm{nm}^{-3}$ in the gas state.
We quench the temperature to $59.8 \mathrm{K}$.
After an equilibration, a liquid-like argon cluster is formed.
We further quench the temperature to $1.2 \mathrm{K}$ to make it rigid, and an amorphous argon cluster is formed \cite{ptp}.
The center of mass of the amorphous argon cluster is placed at $5.1 \mathrm{nm}$ above the center of mass of the graphene.
The argon cluster is translated with the incident velocity $V$ to collide with the graphene.
The incident angle of the argon cluster to the graphene normal is zero.

\section{results
\label{sec:results}}
\subsection{Time evolution of the deflection
\label{sub:observe}}
\begin{figure}
\includegraphics[width = 7 cm]{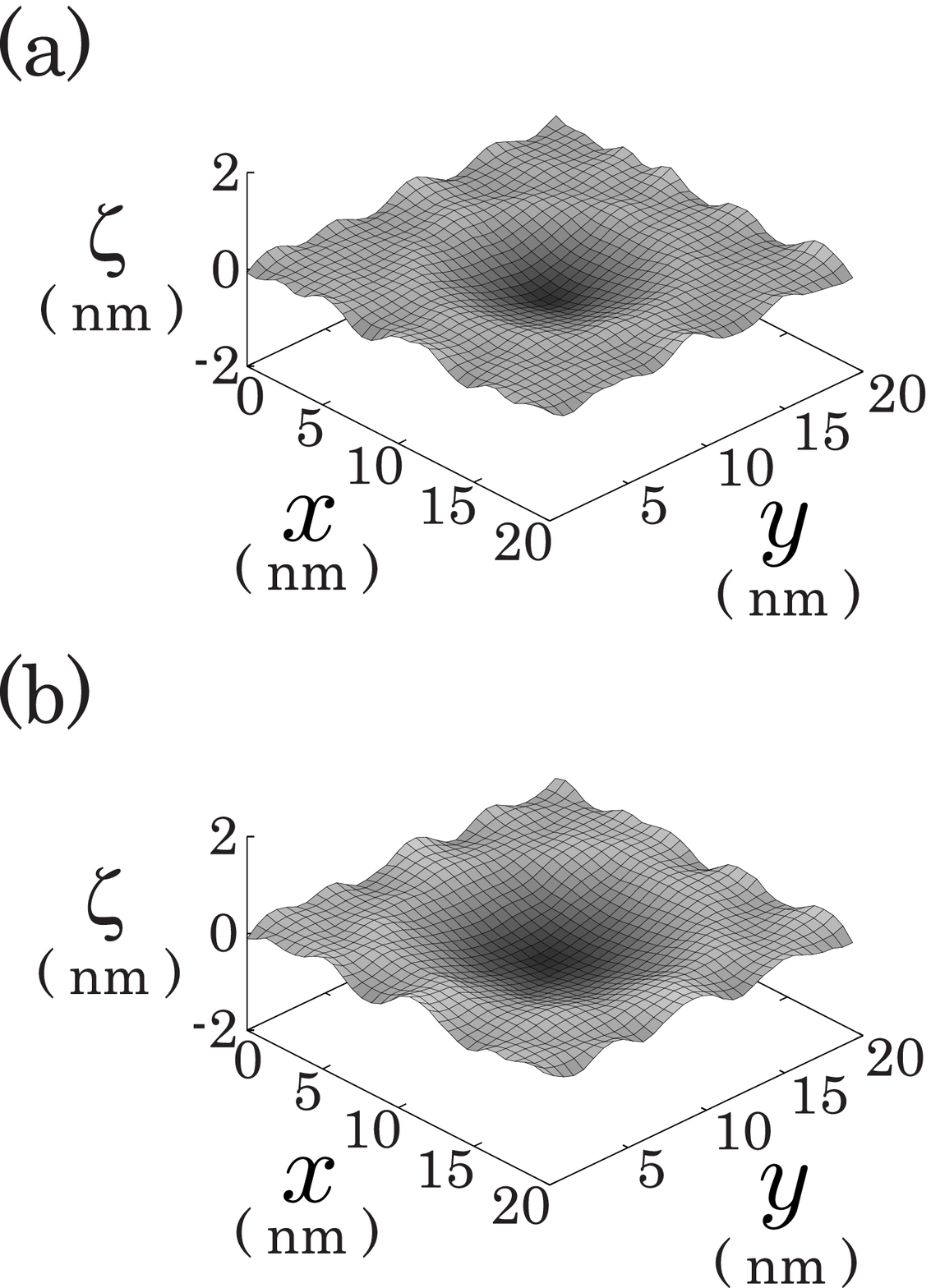}%
\caption{
The deflection of the graphene sheet $\zeta$ at (a) 2.2 ps and (b) 2.8 ps after the initial hitting.
The incident cluster contains $500$ argon atoms, and the incident speed is $316 \mathrm{m/s}$.
\label{fig:view_N500_V2.0}}
\end{figure}
\begin{figure}
\includegraphics[width = 7 cm]{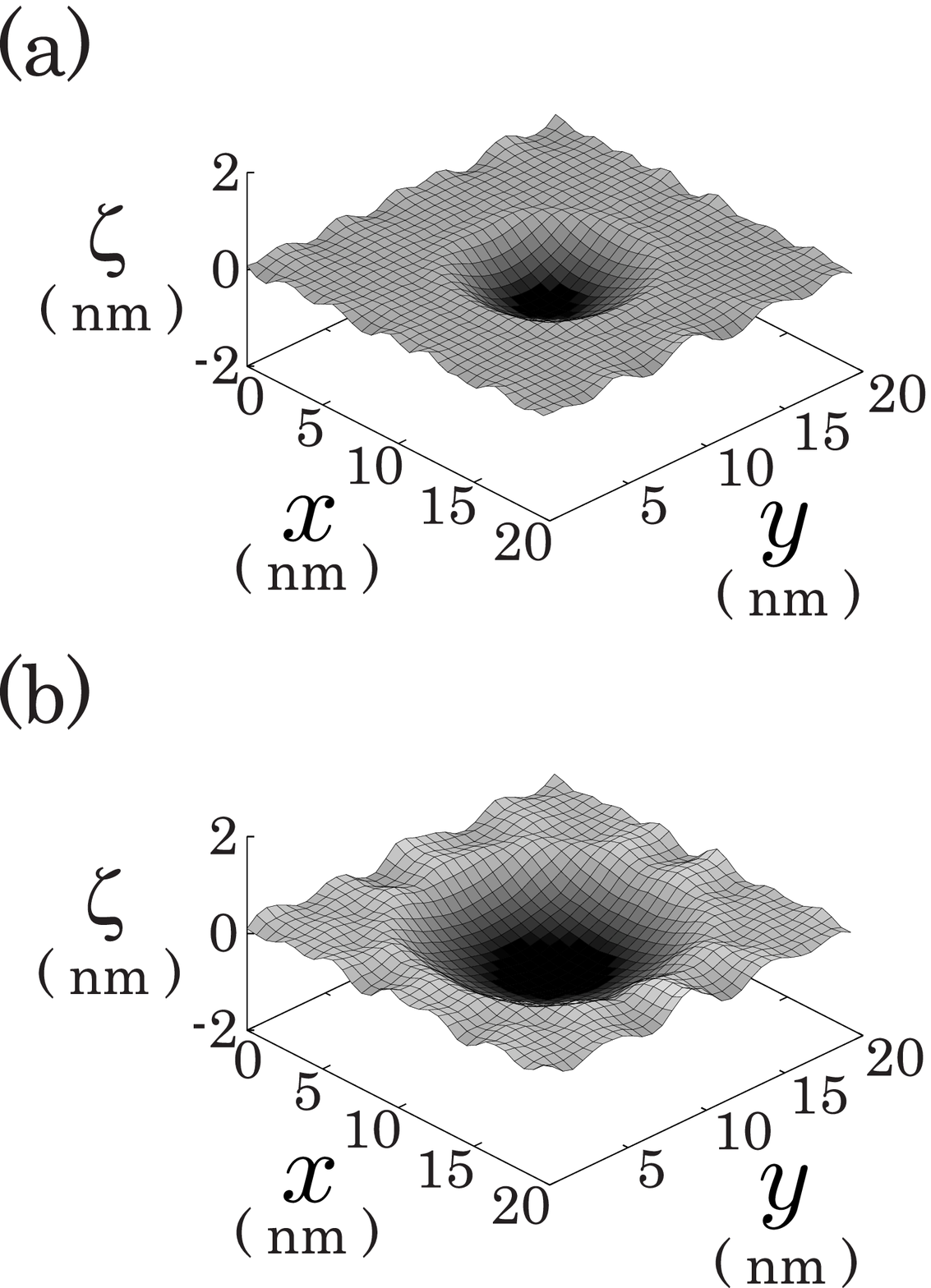}%
\caption{
The deflection of the graphene sheet $\zeta$ at (a) 2.2 ps and (b) 2.8 ps.
The incident cluster contains $500$ argon atoms, and the incident speed is $790 \mathrm{m/s}$.
\label{fig:view_N500_V5.0}}
\end{figure}
Let us demonstrate the motion of the graphene induced by the collision with the argon cluster in the case of $V = 316$~$(\mathrm{m/s})$.
Figures \ref{fig:view_N500_V2.0} display the time evolution of the deflection of the graphene $\zeta$ as a function of $x$ and $y$ coordinates.
In this figures, we divide the $xy$ plane into $32 \times 32$ cells and average over $z$-components of the positions of carbon atoms in the center-of-mass frame.
We define $t=0$ as the time at which the argon cluster contacts the graphene sheet.
At the impact, the circular region around the center of the graphene is bended by the incident argon cluster (~Figs. \ref{fig:view_N500_V2.0}~(a)~),
and the transverse deflection wave is isotropically propagated in the graphene (~Figs. \ref{fig:view_N500_V2.0}~(b)~).
In the laboratory system, the graphene is moved downward and immediately reaches the uniform motion along the $z$-axis with the speed $28.4 \mathrm{m/s}$.
During the impact, the incident argon cluster adsorbs on the graphene and does not rebound.
Figures \ref{fig:view_N500_V5.0} display the time evolution of $\zeta$ for the incident speed $V = 790$~$(\mathrm{m/s})$.
At the impact, the circular region around the center of the graphene is strongly bended by the incident argon cluster (~Figs. \ref{fig:view_N500_V5.0}~(a)~),
and the transverse deflection wave is observed (~Figs. \ref{fig:view_N500_V5.0}~(b)~).
During the impact, the incident argon cluster bursts into fragments and some fragments are scattered and the rest of fragments adsorb on the graphene.
We have also examined the cases of $V = 158$, $474$, and $632$~$(\mathrm{m/s})$,
and the bending formation and the propagation of transverse deflection wave are also observed.
In all cases, the deflection wave in the graphene passes through the boundary without reflection, and the graphene ripples after the impact.
We have never observed any defect formations in the graphene sheet.

\subsection{Analysis of the deflection
\label{sub:deflect}}
\subsubsection{Equation of motion}
To analyze the result of our simulation, we examine the linear theory of the elasticity in description of the deflection of the graphene \cite{elas1,elas2}.
Because the elastic properties of a 2D hexagonal structure are isotropic \cite{elas1}, we ignore the anisotropic properties of the graphene sheet.
Thus, the equation of motion for the deflection is given by
\begin{equation}
\rho \ddot{\zeta}(r,t)+\frac{h^3E}{12(1-\mu^2)}\Delta^2 \zeta(r,t) = p(r,t)~,
\label{eq:motion}
\end{equation}
Here, $\rho = 7.59 \times 10^{-7}$~$(\mathrm{kg/m^2})$ is the mass per unit area of the graphene,
and $\ddot{\zeta}(r,t)$ represents $\partial^2\zeta (r,t)/\partial t^2$.
Because graphene is a single atomic layer of carbon, its thickness $h$ is sometimes set to be the diameter of a carbon atom, $0.335 \mathrm{nm}$.
However, Yakobson \textit{et al.} indicated that $h=0.066$~$(\mathrm{nm})$ should be used in their simulation of single-walled carbon nanotubes \cite{moduli1}.
We still do not have any consensus on the proper value of $h$ \cite{moduli2,moduli3,moduli4,moduli5,moduli6}.
Thus, to avoid ambiguous definition of the thickness, we use the thickness and the elastic moduli which are directly obtained from the analysis of the Brenner potential.
Following Ref.~26, we use the thickness, Young's modulus, and Poisson's ratio as $h=0.0874$~$(\mathrm{nm})$, $E=2.69$~$(\mathrm{TPa})$, and $\mu=0.412$, respectively.
The right hand side of Eq.~(\ref{eq:motion}) is the external pressure due to the argon cluster impact.
Because the deflection is symmetric with respect to the $z$-axis, we assume that $\zeta$ and $p$ depend on time $t$ and the distance from the $z$-axis $r$.
\begin{figure}
\includegraphics[width = 7 cm]{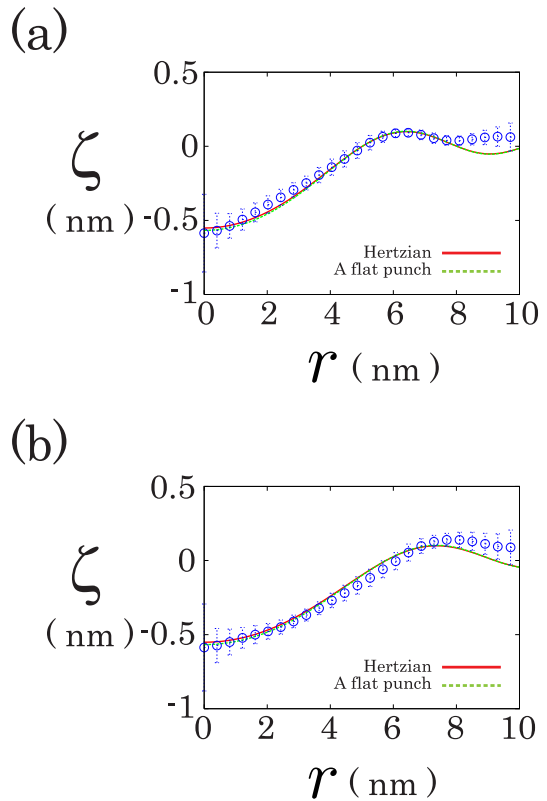}%
\caption{(Color online)
The MD simulation results of the mean deflection of the graphene ( open circle ) which are averaged over the azimuthal coordinate,
and the solutions of the equation of motion, i.e. Eqs.~(\ref{eq:Hertz}) ( red solid line ) and (\ref{eq:punch}) ( green broken line ) at (a) 2.2 ps and (b) 2.8 ps.
The incident cluster contains $500$ argon atoms, and the incident speed is $316 \mathrm{m/s}$.
The magnitude of the impulse is $1.96 \times 10^{-10} \mathrm{N \cdot ps}$.
\label{fig:zetar_N500_V2.0}}
\end{figure}

\subsubsection{Hertzian contact pressure}
Although the external pressure changes during the impact, we simply assume that $p(r,t)$ is an impulsive pressure with the distribution of the Hertzian contact stress.
Thus, we may assume
\begin{equation}
p(r,t)=-\frac{3F}{2\pi a^2}\sqrt{1-\left( \frac{r}{a} \right)^2}\Theta(a-r)\delta(t)~,
\label{eq:pressure}
\end{equation}
where $F$ and $a$ are the impulse and the contact radius of the incident argon cluster, respectively.
Here, $\delta(t)$ is Dirac's delta function, and $\Theta(a-r)$ is the Heaviside function which is defined as $\Theta(a-r)=1$ for $r<a$ and $\Theta(a-r)=0$ for $r>a$.
The contact radius is represented as $\left[ 3FR ( (1-\mu^2)/E + (1-\mu'^2)/E' ) / 4 \right]^{1/3}$ \cite{elas1}.
Here, the mean radius, Young's modulus, and Poisson's ratio of the argon cluster are
$R=1.6$~$(\mathrm{nm})$, $E'=3.69$~$(\mathrm{GPa})$, and $\mu'=0.396$, respectively, which are estimated from our another MD simulation~\cite{kuninaka06,kuninaka09}.
In addition, we assume that the contact area of radius $a$ moves downward with the speed $V$ at the impact.
Thus, the initial conditions of the deflection are $\zeta(r,0)=0$ and $\dot{\zeta}(r,0)=-V\Theta(a-r)$.
Because we consider the behaviors in the vicinity of the center of the graphene, we solve Eq.~(\ref{eq:motion}) as if the graphene sheet is infinitely large.
The Fourier transform and the Laplace transform of Eq.~(\ref{eq:motion}) yield
\begin{equation}
\hat{\zeta}_k(s) = - \frac{H(k)}{s^2+Dk^4}~,
\label{eq:FTPT}
\end{equation}
where $D \equiv h^3E/12\rho(1-\mu^2)$.
Here, we introduce the function
\begin{equation}
H(k) \equiv \frac{3F}{\rho} \frac{ \sin(ak) - ak \cos(ak) }{ (ak)^3 } + 2\pi a V \frac{J_1(ak)}{k}~,
\label{eq:Hk}
\end{equation}
where $J_n(x)$ is the Bessel function for an integer $n$.
Here, we represent the Laplace transform and the Fourier transform as
$\hat{\zeta}_k(s)\equiv  \int_0^\infty \zeta_k(t) e^{-st} dt$ and $\zeta_k(t)\equiv \int_{-\infty}^\infty d\mathbf{k} \zeta(r,t) e^{-i\mathbf{k}\cdot\mathbf{x}}$, respectively.
The inverse Laplace transform and the inverse Fourier transform of Eq.~(\ref{eq:FTPT}) yield
\begin{equation}
\zeta(r,t) = - \int_0^\infty H(k) J_0(kr) \frac{\sin(\sqrt{D}k^2 t)}{\sqrt{D}k} dk~.
\label{eq:Hertz}
\end{equation}

\subsubsection{A flat punch pressure}
If we adopt a flat punch impulsive pressure
\begin{equation}
p(r,t)=-\frac{F}{\pi a^2}\Theta(a-r)\delta(t)
\label{eq:punch_pressure}
\end{equation}
instead of Eq.~(\ref{eq:pressure}), the Fourier transform and the Laplace transform of Eq.~(\ref{eq:motion}) yield
\begin{equation}
\hat{\zeta}_k(s) = - \frac{2H_0J_1(ak)}{ak}\frac{1}{s^2+Dk^4}~,
\label{eq:FL-p}
\end{equation}
where we introduce the constant $H_0 \equiv \rho^{-1}F + \pi a^2 V $.
In this case, the solution of Eq.~(\ref{eq:FL-p}) is
\begin{equation}
\zeta(r,t) = - \frac{2H_0}{a} \int_0^\infty J_0(kr) J_1(ak) \frac{\sin(\sqrt{D}k^2 t)}{\sqrt{D}k^2} dk~.
\label{eq:punch}
\end{equation}

Figures \ref{fig:zetar_N500_V2.0} display the time evolution of the deflection $\zeta(r,t)$ in the case of $V= 316$~$(\mathrm{m/s})$.
In this figures, the open circles are our MD simulation results which are averaged over the azimuthal coordinate.
The red solid and green broken lines represent Eqs.~(\ref{eq:Hertz}) and (\ref{eq:punch}), respectively.
Here, we use $F = 1.96 \times 10^{-10}$~$(\mathrm{N \cdot ps})$ for both Eq.~(\ref{eq:Hertz}) and Eq.~(\ref{eq:punch}).
We have also examined the deflection of the graphene in the case of $V= 158$~$(\mathrm{m/s})$,
and we find that the time evolution of $\zeta(r,t)$ is well described by Eqs.~(\ref{eq:Hertz}) and (\ref{eq:punch}) with $F = 1.25 \times 10^{-10}$~$(\mathrm{N \cdot ps})$.
However, in the cases of $V = 474$, $632$ and $790$~$(\mathrm{m/s})$, Eqs.~(\ref{eq:Hertz}) and (\ref{eq:punch}) are no longer applicable
because the incident argon cluster bursts into many fragments which collide with the graphene,
and the distribution of the external pressure can neither be approximated by the Hertzian contact stress nor a flat punch pressure.

\subsection{Analysis of the heat-up
\label{sub:temp}}
\begin{figure}
\includegraphics[width = 7 cm]{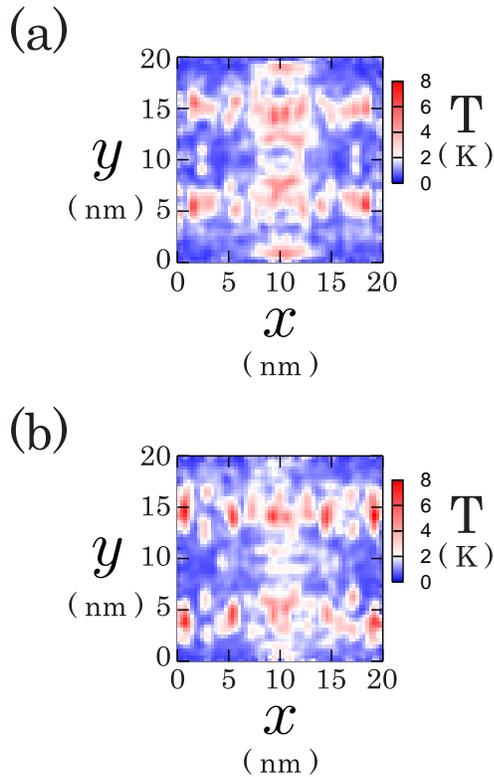}%
\caption{
The temperature profile of the graphene sheet $T(x,y)$ at (a) 2.2 ps and (b) 2.8 ps after the initial hitting.
The incident cluster contains $500$ argon atoms, and the incident speed is $316 \mathrm{m/s}$.
\label{fig:tmp_prof}}
\end{figure}

To study heat up of the graphene, we introduce the local temperature.
We divide the graphene into $64 \times 64$ cells along the $x$- and $y$-axes and define the temperature of the $j$-th cell as
\begin{equation}
T_j = \frac{m}{3k_\mathrm{B}N_j} \sum_{i=1}^{N_j}\left( \mathbf{v}_i - \mathbf{u}_j \right)^2~,
\label{eq:Tj}
\end{equation}
where $k_\mathrm{B}$ and $m=1.99 \times 10^{-26}$~$(\mathrm{kg})$ are the Boltzmann constant and the mass of carbon atom, respectively.
In Eq.~(\ref{eq:Tj}), $N_j$, $\mathbf{v}_i$ and $\mathbf{u}_j$ are the number of carbon atoms in the $j$-th cell,
the velocity of the $i$-th carbon atom which is in the $j$-th cell and the mean velocity of the $j$-th cell, respectively.
The mean velocity of the $j$-th cell is defined as
\begin{equation}
\mathbf{u}_j = \frac{1}{N_j} \sum_{i=1}^{N_j} \mathbf{v}_i~.
\label{eq:meanu}
\end{equation}
In order to take a sample average of $T_j$, we rotate the nanocluster around the line before the impact, where we use the different angle for each sample.
Here, the line is parallel to the $z$-axis and intersects at the center of mass of the nanocluster.
If we project the $64 \times 64$ cells to the $xy$ plane, Eq.~(\ref{eq:Tj}) approximately represents the temperature profile $T(x,y)$.
Figures.~\ref{fig:tmp_prof} display the time evolution of $T(x,y)$ which is averaged over 20 samples in the case of $V=316$~$(\mathrm{m/s})$.
Although the thermal conductivity of a 2D hexagonal structure is isotropic \cite{tcond}, the results of $T(x,y)$ are anisotropic.

Let us explain the anisotropic profile of $T(x,y)$.
The nanocluster collides with the graphene in the vicinity of the center of mass of the graphene $(X,Y)$.
By the impact, the vicinity of $(X,Y)$ is heated up and the heat current $\mathbf{q}$ flows from $(X,Y)$ to the edge of the graphene.
Then, $\mathbf{q}$ is symmetrical with respect to $(X,Y)$, and we adopt $(X,Y)$ for the origin.
In such an irreversible process, thanks to the least dissipation principle, the rate of the entropy production
\begin{equation}
\mathcal{D} = - \int_A \kappa^{-1} \mathbf{q}^2 dA
\label{eq:D}
\end{equation}
is expected to be minimum, where $A$ and $\kappa$ are the area of the graphene and the heat conductivity per unit area of the graphene, respectively~\cite{ldp}.
If we assume that $\kappa$ is a constant, the variation $\delta\mathcal{D}=0$ leads $\mathbf{\nabla}\cdot\mathbf{q}=0$~\cite{ldp}.
Therefore, from Fourier's law of heat conduction, the deviation of the temperature $\delta T = T(x,y) - T_0$ satisfies Laplace's equation $\Delta \delta T = 0$.
Here, $T_0$ is the temperature of the graphene before the impact.
Because $\delta T$ is finite at $(X,Y)$, the general solution of Laplace's equation is
\begin{equation}
\delta T(r,\theta) = const. + \sum_{n=1}^\infty r^n a_n \cos(n\theta+\phi)
\label{eq:T's}
\end{equation}
in the polar coordinate, where $a_n$ and $\phi$ are the integral constants~\cite{PDE}.
Because $\mathbf{q}$ is symmetrical with respect to $(X,Y)$, the integer $n$ satisfies
$\cos\left(n(\theta+\pi)+\phi\right)=\cos(n\theta+\phi)$ and $\sin\left(n(\theta+\pi)+\phi\right)=\sin(n\theta+\phi)$.
Thus, $n$ should be even. Therefore, $\delta T(r,\theta)$ is distributed around $(X,Y)$ as
\begin{equation}
\delta T(r,\theta) = const. + \sum_{m=1}^\infty r^{2m} a_{2m} \cos(2m\theta+\phi)~.
\label{eq:T's2}
\end{equation}
In Fig.~\ref{fig:tmp_prof}~(a), the heated region can be seen as a quadrupole distribution around $(X,Y)$ which is the case of $m=1$ in Eq.~(\ref{eq:T's2}).
On the other hand, in Fig.~\ref{fig:tmp_prof}~(b), the heated region is no longer distributed as Eq.~(\ref{eq:T's2}).
In this case, it seems that the least dissipation principle is no longer correct,
and it is necessary to solve the heat equation with the boundary conditions correctly.

\section{discussion
\label{sec:disc}}
Although Eqs.~(\ref{eq:Hertz}) and (\ref{eq:punch}) seem to well describe the results of our MD simulation, the solution does not satisfy the boundary conditions,
and these are not applicable except for the case that deformations are localized in the vicinity of the center of the graphene sheet, i.e. immediately after the impact.
Note that it is difficult to obtain an analytic solution of Eq.~(\ref{eq:motion}) which satisfies the completely free boundary conditions \cite{fvib1,fvib2}.
If we simply estimate the magnitude of the impulse from the change in momentum of the incident argon cluster,
$F=3.59 \times 10^{-10}$~$(\mathrm{N \cdot ps})$ which is about two times larger than the fitted value in Figs.~\ref{fig:zetar_N500_V2.0}.
However, the value is over-estimated, because the loading force can change during the impact
and the dissipative force plays important role for the collision of clusters \cite{cont1,cont2}.
In the case of $V \ge 400$~$(\mathrm{m/s})$, the impact processes are further complicated by many fragments of the argon cluster,
and Eqs.~(\ref{eq:Hertz}) and (\ref{eq:punch}) are no longer correct.
Therefore, it is necessary to improve the functional form of $p(r,t)$.
Note that if we use $h=0.335$~$(\mathrm{nm})$ in Eqs.~(\ref{eq:Hertz}) and (\ref{eq:punch}),
the wave propagates much faster than the actual propagation observed in our MD simulation.
Thus, the thinner thickness $h=0.0874$~$(\mathrm{nm})$ is more appropriate.
The analysis based on the least dissipation principle reproduces our simulation result of the temperature profile in the early stage of impact.
However, in order to describe the time evolution of the temperature profile, it is necessary to solve the heat equation with appropriate boundary conditions.

\section{conclusion
\label{sec:conc}}
In conclusion, we perform the molecular dynamics simulation of the graphene sheet induced by a collision with an argon nanocluster,
and the bending formation and the propagation of transverse deflection wave are observed.
We find that the linear theory of the elasticity well explains the time evolution of the deflection of the graphene,
where the deflection is represented by using the analytic expressions Eqs.~(\ref{eq:Hertz}) and (\ref{eq:punch}).
In addition, we conclude from the analysis of the motion of the graphene that the actual thickness is much thinner than the diameter of a carbon atom.
We also analyze the time evolution of the temperature profile,
and find that the analysis based on the least dissipation principle reproduces our simulation result in the early stage of impact.
We believe that the predictions of the bending formation and propagation of transverse deflection wave are necessary
for the construction of the nanoscale electronic devices on a graphene sheet.

\begin{acknowledgments}
We thank H. Kuninaka for fruitful discussion and N. Brilliantov for his critical reading of this article.
We also thank the members of department of mathematics in University of Leicester for their hospitality, where parts of this work have been carried out.
This work was supported by the Global COE Program from the Ministry of Education, Culture, Sports, Science and Technology (~MEXT~) of Japan.
This work was also supported by the Research Fellowship of the Japan Society for the Promotion of Science for Young Scientists (~JSPS~),
and the Grant-in-Aid of MEXT (~Grant Nos.~21015016 and 21540384~).
Parts of numerical computation were carried out in computers of YITP, Kyoto University.
\end{acknowledgments}

\bibliography{arsg}

\end{document}